\documentclass[a4paper,10pt,oneside,onecolumn,authoryear]{elsarticle}

\usepackage{array}
\usepackage{graphicx}
\usepackage{lipsum}

\usepackage{url}

\usepackage{amsmath,amssymb}
\usepackage{amsthm}
\usepackage[font=footnotesize]{subfig}
\usepackage{color}
\usepackage{multirow}
\usepackage{rotating}
\usepackage{colortbl}
\usepackage{multirow}
\usepackage{lscape}
\usepackage{longtable}

\biboptions{comma,round}

\newcommand{\goes}[1]{\xrightarrow[]{#1}}

\definecolor{lGray}{rgb}{0.9,0.9,0.9}

\makeatletter
\def\ps@pprintTitle{%
 \let\@oddhead\@empty
 \let\@evenhead\@empty
 \def\@oddfoot{}%
 \let\@evenfoot\@oddfoot}
\makeatother

\begin{document}

\begin{frontmatter}



\title{Estimation and Modelling of PCBs Bioaccumulation in the Adriatic Sea Ecosystem}

\author[l1,l5]{Marianna Taffi}
\author[l2]{Nicola Paoletti}
\author[l3]{Pietro Li\`{o}}
\author[l4]{Luca Tesei}
\author[l1]{Sandra Pucciarelli}
\author[l5]{Mauro Marini}
\address[l1]{School of Biosciences and Veterinary Medicine, University of Camerino, Via Gentile III Da Varano 62032, Camerino, Italy}
\address[l2]{Department of Computer Science, University of Oxford, Parks road, Oxford OX1 3QD, United Kingdom}
\address[l3]{Computer Laboratory, University of Cambridge, 15 JJ Thomson Ave, Cambridge CB3 0FD, United Kingdom}
\address[l4]{School of Science and Technology, Computer Science division, via Del Bastione 1, 62032 Camerino, Italy}
\address[l5]{National Research Council, Institute of Marine Science (ISMAR), Largo Fiera della Pesca 2, 60125 Ancona, Italy}

\begin{abstract}
Persistent Organic Pollutants represent a global ecological concern due to their ability to accumulate in organisms and to spread species-by-species via feeding connections. 
In this work we focus on the estimation and simulation of the bioaccumulation dynamics of persistent pollutants in the marine ecosystem, and we apply the approach for reconstructing a model of PCBs bioaccumulation in the Adriatic sea, estimated after an extensive review of trophic and PCBs concentration data on Adriatic species. 
Our estimations evidence the occurrence of PCBs biomagnification in the Adriatic food web, together with a strong dependence of bioaccumulation on trophic dynamics and external factors like fishing activity. 
\end{abstract}


\end{frontmatter}



\section{Introduction}
Chemical contamination is one of the strongest and most complex abiotic perturbations threatening the stability of marine ecosystems. Constantly, an unrestrained flow of pollutants is released into the seas triggering unpredictable ecological consequences in marine biota, having also cascade effects in the entire ecosystem. The aquatic environment is characterized by multiple contamination pathways making marine organisms particularly prone to bioaccumulation and biomagnification phenomena (defined in Table \ref{tbl:definitions}), of which
Persistent Organic Pollutants (POPs) are globally recognised as one of the main and worst causes.

Chemically belonging to the same class of POPs, Polychlorinated biphenyl (PCBs) have been for a long time matter of ecotoxicological interest for their property of binding with the fatty tissue of animals and, thus, of spreading through feeding connections amplifying their toxic perturbations species-by-species. As a consequence, feeding links become a critical medium of chemical transport across trophic levels up to the human being \citep{kelly2007food,lohmann2007global}, which makes the entire marine ecosystem both sink and source of hazardous substances.

Bioaccumulation represents a complex and exhaustive ecotoxicological indicator to evaluate the toxic exposure of living organisms. Its definition \citep{mackay2000bioaccumulation} 
includes abiotic and biotic factors (see Table~\ref{tbl:definitions}), therefore it can vary for the same species surveyed in different environments. Variations may depend on the multiple patterns of uptake and their related concentrations, on the nature of xenobiotics, and on species-specific biological characteristics (lipid content, size, age and etc). The combination of all these aspects with the variety of local environmental and ecological scenarios makes difficult to obtain an all-encompassing model. 

In the field of environmental toxicology, prediction-oriented works are still relatively recent. However, over the last decades this area has considerably evolved with advances in models, approaches and indices to assess chemicals fate and effects on species and communities \citep{van2003fish,mackay2000bioaccumulation, arnot2006review, devillers2009ecotoxicology, jorgensen2001fundamentals}. What makes a model exploitable for predictive purposes is its practical utility and, above all, its ability to reproduce bioaccumulation dynamics in a robust way \citep{luoma2005metal}.

As opposed to regression-based approaches for predicting species bioconcentration or bioaccumulation, mechanistic models \citep{mackay2000bioaccumulation, nichols2009bioaccumulation} require specifying and quantifying the different pathways of contaminant uptake and loss. Thus, they provide a more detailed characterization of the bioaccumulation dynamics and can accommodate the use of empirical data so that estimations and predictions are able to reflect measured conditions. In addition, toxicokinetic-toxicodynamic (TKTD) models study the effects of specific molecules at finer scales, including the diffusion within internal compartments (fat tissue, liver, muscle and etc) and the responses at the genetic and cellular levels. It is clear that a large amount of high quality, species-specific, toxicity data are needed to parametrize TKTD models. Therefore, data availability mainly drives the choice of a particular model and its degree of detail. Besides this, simpler models could provide, with less effort, a fair accuracy or predictive power and having more specific models do not necessarily imply better results \citep{wainwright2013environmental, wania1999evolution}.

In this work, we derive a \textit{bioaccumulation model of PCBs in the Adriatic sea}, based on one of the most complete trophic studies in this area \citep{coll2007ecological}. Since PCBs necessarily follow feeding links, we take a mechanistic approach where pathways of contaminant uptake and loss correspond to trophic flows among species and environmental compartments. What makes the Adriatic area of crucial interest from an ecotoxicological point of view is the combination of high environmental variability and biodiversity with several anthropogenic factors, and a peculiar geography and orography.

A number of studies on PCBs concentration in Adriatic species have been carried out during the last decades. Using this literature we have reconstructed a database that includes toxicity data for each considered species (see Table~\ref{tbl:PCBinputdata}). However, due to the lack of data for some compartments and their temporal and spatial patchiness, the resulting collection is still incomplete. Hence, we use Linear Inverse Modelling (LIM), a technique typically employed in trophic estimation, to calculate the missing information. In particular, we obtain, as a first result, a static bioaccumulation network model from which we parametrize a Lotka-Volterra ODE system used to analyse the contaminant long-term dynamics through the food web.

With our methodological framework, we show the occurrence of biomagnification in Adriatic species and that bioaccumulation mainly depends on trophic aspects (e.g.\ diet and fishing).

\begin{table}
\centering
\caption{Glossary of ecotoxicological concepts.}\label{tbl:definitions}
\begin{tabular}{p{0.95\columnwidth}}
\hline
\textit{\textbf{Bioconcentration}} 
is the net process of chemical adsorption only from water through respiration and dermal surfaces minus eliminations routes (respiration, egestion, growth dilution and metabolic tranformation).\\

\textit{\textbf{Bioaccumulation}} refers to the contamination process of an organism resulting from all possible paths of uptake: waterborne (bioconcentration) and dietary sources, net of chemical elimination routes.\\ 
\textit{\textbf{Biomagnification}} is the phenomenon by which predators have higher concentrations than their prey, leading to increasing accumulation of pollutants along increasing trophic levels.\\
\textit{\textbf{Polychlorinated Biphenyls}} are synthetic chemical compounds, chemically defined as chlorinated hydrocarbons. They are highly stable molecules with general formula $C_{12} H_{(10-n)} Cl_n$   ($n=1,\ldots,10$ number of chlorine atoms), consisting in a group of 209 different congeners.
Congeners vary in the degree of chlorination and chlorine position (para, meta and ortho).
Higher chlorine contents correspond to higher environmental persistence and lower biodegradability (photolytic, biological and chemical).
PCBs are characterized by semi-volatility, low water solubility, low vapour pressure and long-range transport. PCBs have been used in hundreds of commercial and industrial application until their ban in the 80's, but are still ubiquitous and have been traced even much farther than their application point, like in Arctic regions~\citep{norstrom1988organochlorine}.
Biologically, they are highly noxious for aquatic living organisms for their \textit{lipophilicity}, i.e.\ the ability to dissolve in fats and lipids (including oils and non-polar solvents). This gives rise to the phenomena of \textit{bioaccumulation} and \textit{biomagnification} in marine food webs, up to the human being \citep{dewailly1989high}. 
\\
\hline
\end{tabular}
\end{table}
\section{Materials and Methods}

\subsection{The Adriatic Sea Ecosystem}
This study focuses on the Adriatic Sea, a relatively shallow sub-basin of the Mediaterranean Sea (max depth 1250 m) with limited extension (800 km major axis - 200 km minor axis) but characterized by high ecological relevance and environmental variability \citep{cushman2001physical}, wide diversity of marine species and microbial communities \citep{coll2010biodiversity}.

This area is exposed to multiple external forcing mechanisms that combined, lead to adverse effects in the pollution load spilled into the Adriatic ecosystem. The Adriatic region  is characterized by high anthropogenic activities \citep{de2004geochemistry} and river discharge fluctuations. 

Geographically, the complex orography of the gulf of Trieste and the Venice lagoon causes a significant penetration and exchange of coastal water into the urban environment. In particular the Po river by crossing a wide industrial and agricultural area, represents the major buoyancy input with an annual mean discharge rate of 1500 $m^{3} \cdot s^{-1}$ accounting for the third of the total riverine freshwater input into the Adriatic Sea Sea \citep{campanelli2011influence}. Moreover, the Southern Eastern Adriatic rivers are equally an important potential source of pollutants, being the mainly entrance to the Southern part of the Adriatic Sea \citep{marini2010southeastern}.
In relation to this aspects, several surveys conducted in different Adriatic regions report significant concentrations of different xenobiotics detected both in species and environmental compartment \citep{marini2012evaluation,bellucci2002distribution,horvat1999mercury,kannan2002perfluorooctanesulfonate}

\subsection{Input data}
In order to define the PCBs bioaccumulation model, we firstly need to estimate the trophic network of Adriatic species. We start from data reported in \citep{coll2007ecological}, one of the most complete quantitative study of the Northern and Central Adriatic food web in which forty functional groups have been identified to investigate the ecological impact of fishing activity during the decade 1990-2000. 
In their work the Adriatic trophic model is developed with ECOPATH \citep{christensen2004ecopath}, and estimated through a mass balanced approach by using literature and survey data on species abundance.
In our model, we follow the same functional group classification as in \citep{coll2007ecological}, taking the same input data (reported in Table \ref{tbl:biomassInOut}) regarding biomass ($B$, measured in $t\cdot km^{-2}$ wet weight organic matter); production/biomass ratio ($P$, i.e.\ rate of biomass generation, $yr^{-1}$) and consumption/biomass ratio ($Q$, i.e.\ rate of biomass losses, $yr^{-1}$). Data on fishing activity has also been taken into account, but not reported here. Diet composition is illustrated in Fig.~\ref{fig:dietcompositions} (a) of the supplementary material. Biomass flows are expressed in $t\cdot km^{-2} \cdot yr^{-1}$. 

We reviewed a large amount of literature dealing with field analysis of PCBs bioconcentration in marine species of the Adriatic Sea conducted over the last decades. 
In order to maintain homogeneity with trophic data, we follow some criteria in the selection of the available PCB concentration data.
In particular, we collected data sampled in species in North, Central and South Adriatic area during the period 1994-2002, where PCBs concentrations in marine organisms have been determined in edible parts and muscle tissue; and, as usually done in field surveys, we considered the sum of PCBs congeners expressed in $ng \cdot g^{-1}$ wet weight-based.
When PCBs values are not available for the same species identified in \citep{coll2007ecological}, we select concentration data of Adriatic species with the same taxonomic classification.
 Table \ref{tbl:PCBinputdata} summarizes the PCBs concentration data used in the estimation of the bioaccumulation model for each functional group, together with the corresponding sampled species and literature reference. Details on the sampling period, geographic area, tissue analysed and PCBs congeners detected are summarized in Tab. \ref{tbl:detailedPCBrefs} of the supplement.

\subsection{Food web estimation with Linear Inverse Modelling}
Following \citep{ulanowicz2004quantitative}, to describe an ecological network we need its \textit{topology} (qualitative information), i.e.\ the nodes representing the relevant groups and the directed edges representing the feeding links; and its \textit{flow rates} (quantitative information), i.e.\ for each edge, the rate at which a medium (in our case, biomass or contaminant) is transferred from the source (the prey) to the target (the predator).
Generally, flow rates are estimated at some equilibrium conditions, according to which functional groups are mass-balanced, i.e.\ the inflows must equal the outflows.
In addition, each group possesses an internal storage of biomass/bioconcentration affecting the value of flow rates. Food web models need to include also external compartments, used to implement exogenous imports and exports. Externals are not mass-balanced, thus they represent potentially unlimited sources and sinks of medium.

In the following, we denote with $b_{i\goes{}j}$ and $c_{i\goes{}j}$ the flow rate of biomass and contaminant, respectively, from compartment $i$ (the prey) to $j$ (the predator); and the storages $B_i$ and $C_i$ are used to indicate the biomass and PCBs concentration of $i$, respectively.

One of the most extensively used techniques in reconstructing feeding connections from empirical data is \textit{Linear Inverse Modelling (LIM)}, through which the food web is described as a linear function of its unknown flows. The term \textit{inverse} indicates that such unknown flows are determined from empirical data, put in the model by means of linear equalities and inequalities \citep{van2010quantifying}. A LIM problem can be formulated as
\begin{align}
min \ &\ |A\cdot \mathbf{x} - \mathbf{b}|^2\label{eq:LIM0}\\
subject\ to \ &\ A\cdot \mathbf{x} \simeq \mathbf{b}\label{eq:LIM1}\\
&\ E\cdot \mathbf{x} = \mathbf{f}\label{eq:LIM2}\\
&\ G\cdot \mathbf{x} \geq \mathbf{h},\label{eq:LIM3}
\end{align}
where $\mathbf{x}$ is the vector of unknown flows; Eq.~\ref{eq:LIM1} indicates an optional set of equalities that are approximately met (as closest as possible); the strict equalities in Eq.~\ref{eq:LIM2} are used to model hard constraints, typically to incorporate mass balances and high quality data; and inequalities in~\ref{eq:LIM3} are used to models soft constraints (e.g.\ when dealing with low quality data). Among the different methods available for solving a LIM problem, in this work we use the \textit{least square method} that attempts to minimize the squared difference between the estimates and the data in the approximate equations (see Eq.~\ref{eq:LIM0}), and solutions are accepted up to a fixed tolerance value.

When approximate equalities are excluded and the solution space is not-empty (i.e.\ when dealing with an \textit{under-determined} model), a single solution can be picked up that minimizes the sum of squared flows or other global ecosystem properties expressible as linear functions. Alternatively, the solution space can be explored through Monte-Carlo sampling for finding the flows that are most likely under a statistical viewpoint. Then, the mean of the sampled solutions can be taken as a single (valid) solution, as shown in \citep{van2010quantifying}.

\subsection{Adriatic PCBs bioaccumulation model}
\subsubsection{Conceptual model}
\begin{figure*}
\centering
\includegraphics[trim=0 3.9cm 0 2.2cm, clip, width=0.9\linewidth]{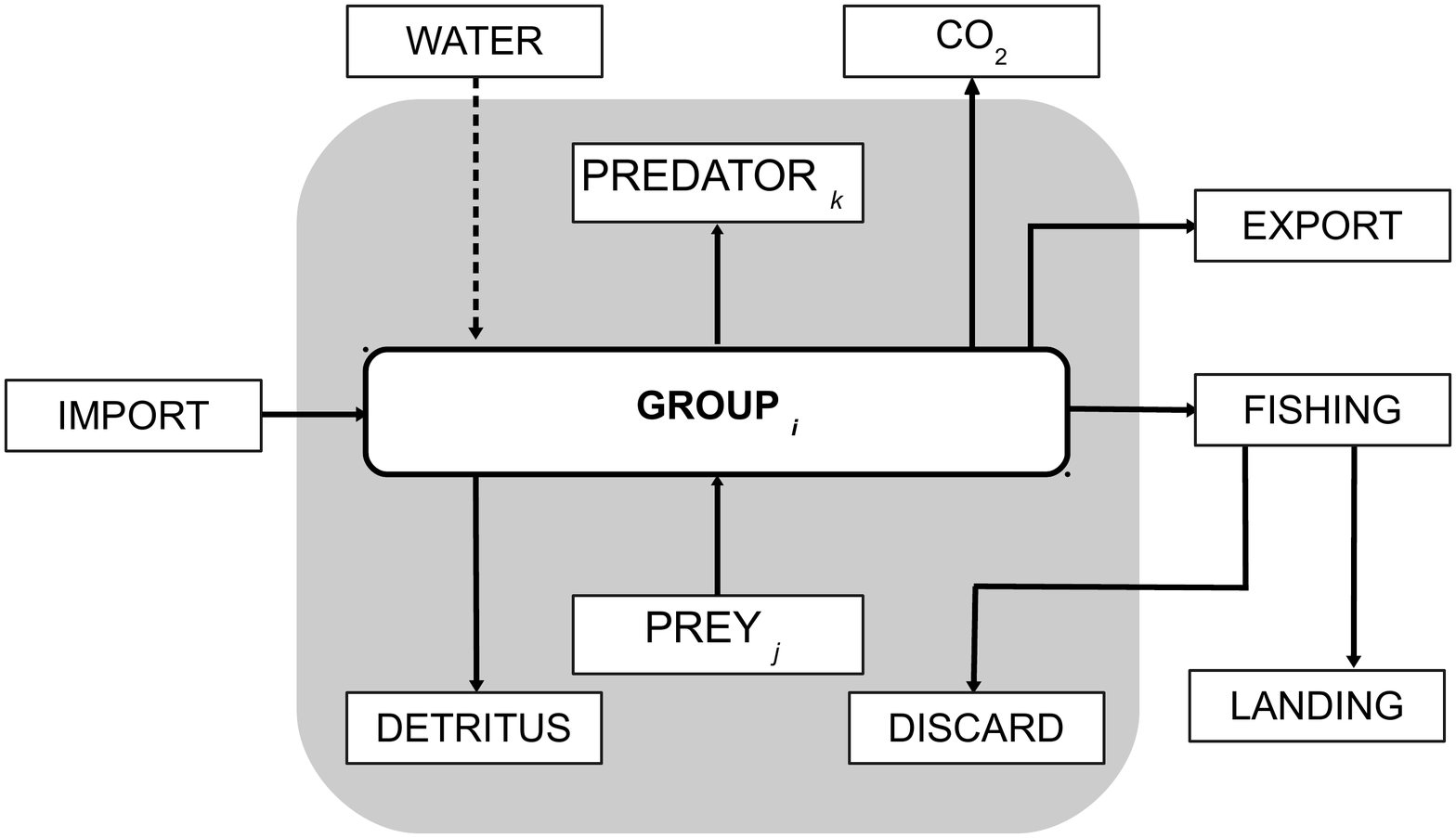}
\caption{Conceptual model of the Adriatic PCBs bioaccumulation network. Mass-balanced groups are enclosed in the gray box; externals are depicted outside of the box. Generic compartments $i$, $j$, $k$ correspond to any of the mass-balanced groups. Solid arrows indicate combined biomass and contaminant flows between groups. The dashed arrow from the water compartment represents the only contaminant uptake flow which is not mediated by biomass.}\label{fig:conceptual}
\end{figure*}

Figure \ref{fig:conceptual} illustrates the conceptual model describing the biomass and the PCBs flows between the groups of our network. Given a generic (mass-balanced) group $i$, we consider:
\begin{itemize}
\item consumption ($b_{j\goes{}i}$) or contaminant uptake ($c_{j\goes{}i}$) from a prey $j$;
\item predation ($b_{i\goes{}k}$) or contaminant losses ($c_{i\goes{}k}$) due to consumption by a predator $k$;
\item external inputs of biomass and PCBs from $Import$ ($b_{Import\goes{}i}$ and $c_{Import\goes{}i}$), used to model generic inflows like immigration;
\item flows describing contaminant uptake from water ($c_{Water\goes{}i}$), which clearly do not involve a corresponding biomass transfer;
\item external outputs to $Export$ ($b_{i\goes{}Export}$ and $c_{i\goes{}Export}$), describing generic outflows like emigration;
\item outflows to the $CO_2$ compartment ($b_{i\goes{}CO_2}$ and $c_{i\goes{}CO_2}$), modelling respiration flows and that together with the flows to detritus account for the unassimilated portion of ``ingested'' food; and
\item outflows due to fishing activity, which can be directed to the landings ($b_{i\goes{}Landing}$ and $c_{i\goes{}Landing}$) or to the discards ($b_{i\goes{}Discard}$ and $c_{i\goes{}Discard}$). The latter enters back the biomass cycle and is modelled as a mass-balanced group.
\end{itemize}

\subsubsection{Trophic Network} \label{sec:trophic}
Table~\ref{tbl:trophic_lim} summarizes the linear equalities and inequalities defined for the estimation of biomass flows.
The results of the quantification are discussed in Section~\ref{sect:res_net} (Fig. \ref{fig:estimated_net} (a) gives a graphical illustration of the estimated network and Table \ref{tbl:biomassInOut} reports the numerical results).

\begin{table}
\caption{Linear Inverse Model for the estimation of trophic flows. $b_{i\rightarrow j}$ denotes the biomass flow from group $i$ to group $j$.}
\begin{small}
\centering
\renewcommand{\arraystretch}{1.15}
\begin{tabular}{p{0.95\columnwidth}}
\hline
\textbf{Mass balances:}$\quad \frac{db_i}{dt} = \sum_j b_{j\rightarrow i} - \sum_j b_{i\rightarrow j} = 0$\\
The difference between inflows and outflows is zero for each functional group $i$; $j$ ranges among groups and externals.\\ \hline 
\textbf{Ingestion$^1$:}$\quad I_i = \sum_{j} b_{j\rightarrow i} = B_i \cdot Q_i$\\ 
The total ingestion $I_i$ of a group $i$, i.e.\ the sum of all the consumption flows, equals the product between biomass $B_i$ and consumption rate $Q_i$; $i$ and $j$ are functional groups; for $i$, we exclude detritus and primary producers (phytoplankton).\\ \hline
\textbf{Unassimilated Food:}$ \quad b_{i\rightarrow CO2} + b_{i\rightarrow Detritus} = I_i \cdot (1 - g_i)$\\
Respiration flows $b_{i\rightarrow CO2}$ and flows to detritus $b_{i\rightarrow Detritus}$ constitute together a fraction of the total ingestion and accounts for the proportion of food that is not converted into biomass. $g_i = \frac{P_i}{Q_i}$ is the gross food conversion efficiency~\citep{christensen2004ecopath}. \\  \hline
\textbf{Respiration-assimilation:}$\quad b_{i\rightarrow CO2} \leq I_i \cdot g_i$\\
As pointed out in \citep{coll2007ecological}, the ratio respiration/assimilation has to be lower than one, in order to have realistic estimates.\\  \hline
\textbf{Diet$^1$:}$\quad b_{i \rightarrow j} = I_j \cdot DC_{ij}.$\\
The biomass flow from prey $i$ to predator $j$ is given by the proportion of the total ingestion of $j$ coming from $i$. $DC$ is the diet composition matrix (Fig. \ref{fig:dietcompositions} (a)).\\ \hline
\textbf{Non-negativity of flows:}$ \quad b \geq 0$\\ \hline
\multicolumn{1}{p{0.95\columnwidth}}{\footnotesize $^1$For species with uncertain input biomass and diet data, appropriate inequalities are set.}
\end{tabular}
\end{small}
\label{tbl:trophic_lim}
\end{table}

\subsubsection{Bioaccumulation Network}\label{sect:bioacc_net}
\begin{table}
\caption{Linear Inverse Model for the estimation of PCB concentrations. $c_{i\rightarrow j}$ denotes the contaminant flow from group $i$ to group $j$.}\label{tbl:toxic_lim}
\centering
\begin{small}
\renewcommand{\arraystretch}{1.15}
\begin{tabular}{p{0.95\columnwidth}}
\hline
\textbf{Mass balances:}$\quad \frac{dc_i}{dt} = \sum_j c_{j\rightarrow i} - \sum_j c_{i\rightarrow j} = 0$\\
For each group $i \in KINE$, bioconcentrations are estimated under the mass-balance assumption; $j$ ranges among groups and externals.\\ \hline
\textbf{Concentration data:}$\quad C_i = PCB_i$\\
These equations incorporate PCB input data ($PCB_i$) into the model. Inequalities are used in correspondence of groups with uncertain input PCB concentrations. $i \in KINE \cup INST$.\\ \hline
\textbf{Uptake from food/losses:}$\quad c_{j\rightarrow i} = b_{j \rightarrow i} \cdot C_j$\\
The contaminant flow from $j \in KINE \cup INST$ to $i$ is the product of the corresponding biomass flow $b_{j \rightarrow i}$ and the PCB concentration in $j$. If $i \in KINE$, the equation describes the uptake from food by predator $i$. If instead $i$ is an external, $c_{j\rightarrow i}$ represents a contaminant loss by $j$.
\\ \hline
\textbf{Uptake from generic imports:}$\quad c_{import \rightarrow i} = b_{import \rightarrow i} \cdot C_i$\\
The concentration of the biomass imported into group $i \in KINE$) is assumed to be the same as in $i$.\\  \hline
\textbf{Uptake from environment:}$\quad c_{water \rightarrow i} = w_i  \cdot C_{water}$\\
$w_i$ is the rate of contaminant uptake from water by group $i \in KINE$ and $C_{water}$ is the concentration in water$^{1}$.\\  \hline
\textbf{Non-negativity of concentrations:}$ \quad C_i \geq 0$\\ \hline
\multicolumn{1}{p{0.95\columnwidth}}{\footnotesize $^{1}w_i$ cannot directly estimated, since it depends on a non-linear constraint ($w_i$ and $C_{water}$ are both unknowns). $c_{water \rightarrow i}$ is instead computed and $w_i$ calculated subsequently.}
\end{tabular}
\end{small}
\end{table}

Following~\citep{hendriks2001power,laender2009incorporating}, our bioaccumulation model distinguishes between two kinds of functional groups:
\begin{description}
\item[$i \in INST$:] compartments modelling small particles that are assumed to be in instant equilibrium with the water phase, like detritus and planktonic groups. For such groups, contaminant concentration is computed as:
\begin{equation*}
C_i = C_{water}\cdot OC_{i} \cdot K_{OC}
\end{equation*}
where $C_i$ is the concentration of $i$ and $OC_i$ is its organic carbon fraction; $C_{water}$ ($\mu g/L$) is the unknown concentration in water and $K_{OC}=k_{OC/OW}\cdot K_{OW}$ is the organic carbon-water partition ratio, calculated as a function of $K_{OW}$, the octanol-water partition ratio of the contaminant. $K_{OW}$ is a measure of how a compound is hydrophilic or hydrophobic, and in this case its value depends on the particular PCB congener considered. Since we consider the sum of congeners, we set $K_{OW}=10^6$, given that the $Log K_{OW}$ of PCBs varies between 5 and 7, as reported in~\citep{walters2011trophic}. The other parameters have been taken from~\citep{laender2009incorporating}: $OC_i = 0.028$ and $k_{OC/OW}=0.41$.
\item[$i \in KINE$:] compartments whose concentration depends on the amount of contaminant exchanged through biomass flows, and are estimated as in Table \ref{tbl:toxic_lim}. This option applies to groups where contaminant uptake and losses resulting from absorption, ingestion, egestion, excretion and growth cannot be neglected \citep{hendriks2001power}.
\end{description}

We formulate a linear inverse model (Table~\ref{tbl:toxic_lim}) where PCB concentrations $C_i$ are the unknowns to estimate, differently from the trophic model where biomass flows are the unknown variables. Contaminant flows are expressed in $ng \cdot g^{-1} \cdot t \cdot km^{-2} \cdot y^{-1}$.
In Section~\ref{sect:res_net} we illustrate the estimated PCB bioaccumulation network and we evaluate biomagnification phenomena on the species of the Adriatic ecosystem. Numerical results are reported in Table \ref{tbl:pcb_est}.

\subsection{Derivation of ODE Model}\label{sect:ode_model}
Recalling that PCB diffusion follows the same paths as biomass flows determined by prey-predator relationships, we define a dynamic bioaccumulation model on top of a multi-species Lotka-Volterra system used to describe the temporal changes in species biomass. In its general form, the system is formulated as:
\begin{equation}\label{eq:lv}
\dot{B}_i(t) = B_i(t)\left(g_i - \sum_j A_{ij} B_j(t)\right)
\end{equation}
where $B_i(t)$ is the biomass of species $i$ at time $t$; $g_i$ is the intrinsic growth rate of $i$; and $A$ is the interaction matrix describing inter-specific effects. In particular $A_{ij}$ describes the predation effect of species $j$ on species $i$.
Although there are several possible ways to derive population-dynamics parameters from a food web model \citep{palamara2011population}, here we follow quite a standard approach:
\begin{itemize}
\item $B_i(0)=B_i$, initial biomass values are those in the static food web estimated with LIM;
\item $g_i = \dfrac{\sum_j b_{j \rightarrow i} - \sum_j b_{i \rightarrow j}}{B_i(0)}$, with $j$ ranging among the external groups: the growth rate of $i$ is the sum of exogenous inflows and outflow, over the estimated biomass of $i$;
\item $A_{ij} = \dfrac{b_{i\rightarrow j} - b_{j\rightarrow i}}{B_i(0) \cdot B_j(0)}$, the interaction rate between prey $i$ and predator $j$ is calculated as the net flow going from $i$ to $j$ divided by the estimated biomasses of $i$ and $j$. 
\end{itemize}
Additionally, we can define the biomass flow rate from group $i$ to $j$ at time $t$, $b_{i\rightarrow j}(t)$, which is non-linear with respect to the biomasses of $i$ and $j$, as:
$$b_{i\rightarrow j}(t) = \dfrac{b_{i\rightarrow j}}{B_i(0) \cdot B_j(0)} \cdot B_i(t) \cdot B_j(t)$$
in a way that Eq. \ref{eq:lv} can be rewritten as:
$$\dot{B}_i(t) = g_i \cdot B_i(t) + \sum_j b_{j\rightarrow i}(t) - \sum_j b_{i\rightarrow j}(t)$$
Therefore, the dynamics of the contaminant concentration in species $i$, $C_i(t)$, is given by the net sum of contaminant flows, over the biomass of $i$:
\begin{multline}\label{eq:lv_cont_1}
\dot{C}_i(t) = w_i \cdot C_{water} + \\ \dfrac{g_i \cdot B_i(t) \cdot C_i(t) + \sum_j b_{j\rightarrow i}(t)\cdot C_j(t) - \sum_j b_{i\rightarrow j}(t) \cdot C_i(t)}{B_i(t)}
\end{multline}
where $C_{water}$ is the concentration in water (assumed constant) and $w_i$ is the uptake rate from water by group $i$. As done for the biomass equations, the initial concentrations correspond to those estimated in the static bioaccumulation network: $C_i(0)=C_i$, for each group $i$.

Finally, expanding the interaction terms, Eq. \ref{eq:lv_cont_1} is equivalent to the following:
\begin{multline}\label{eq:lv_cont_2}
\dot{C}_i(t) = w_i \cdot C_{water} + g_i \cdot C_i(t) + \sum_j  \left( \dfrac{b_{j\rightarrow i}}{B_i(0) \cdot B_j(0)} \cdot B_j(t) \cdot C_j(t) \right) \\ - \sum_j \left( \dfrac{b_{i\rightarrow j}}{B_i(0) \cdot B_j(0)}  \cdot B_j(t) \cdot C_i(t) \right)
\end{multline}

In the remainder of the paper, we will focus on the temporal changes in bioconcentrations independently of the biomass variations, thus assuming constant species biomass ($B_i(t) = B_i(0), \forall t$), which gives the following system of linear differential equations:
\begin{equation}\label{eq:lv_cont_3}
\dot{C}_i(t) = w_i \cdot C_{water} + g_i \cdot C_i(t) + \sum_j  \left( \dfrac{b_{j\rightarrow i}\cdot C_j(t) - b_{i\rightarrow j} \cdot C_i(t)}{B_i(0)} \right)
\end{equation}
Note that this simplification does not change the quantitative dynamics of the model, because biomasses have been estimated under mass-balance conditions.
\section{Results and Discussion}
\subsection{Estimated Trophic and Contaminant Network}\label{sect:res_net}
\begin{figure*}
\centering
\subfloat[Biomass network]{\includegraphics[width=\linewidth]{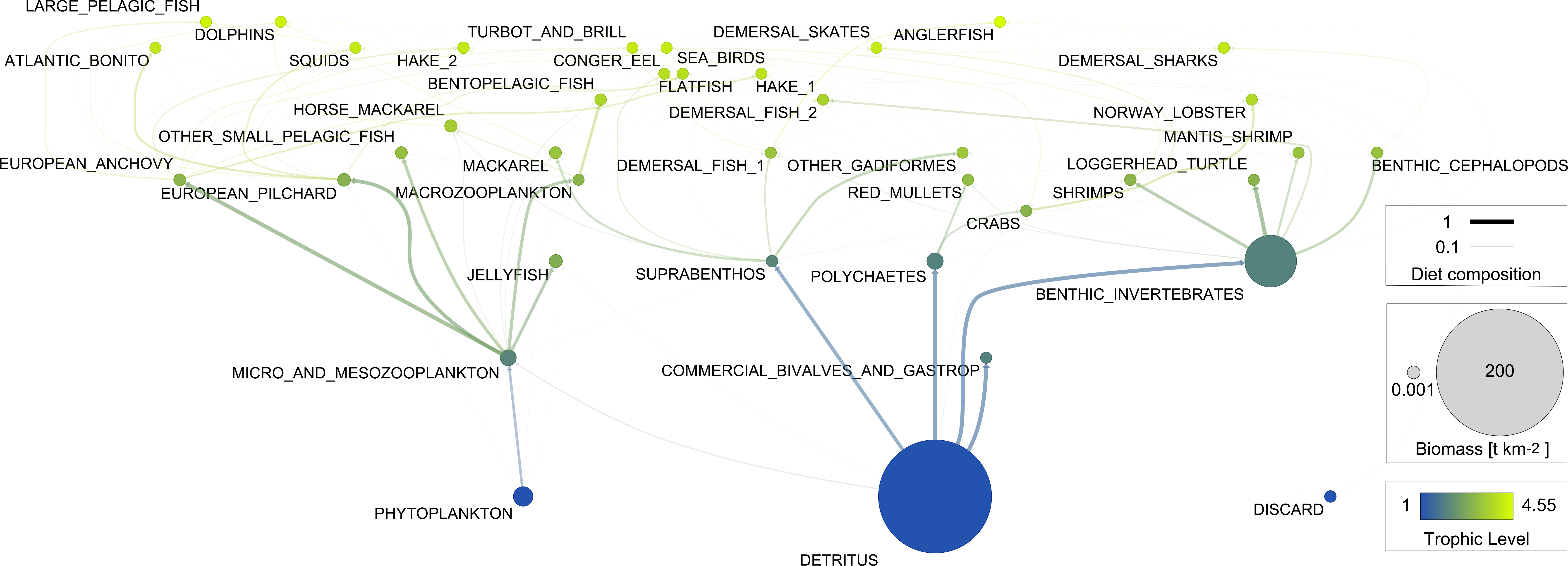}}\\ \vspace*{1cm}
\subfloat[Bioaccumulation network]{\includegraphics[width=\linewidth]{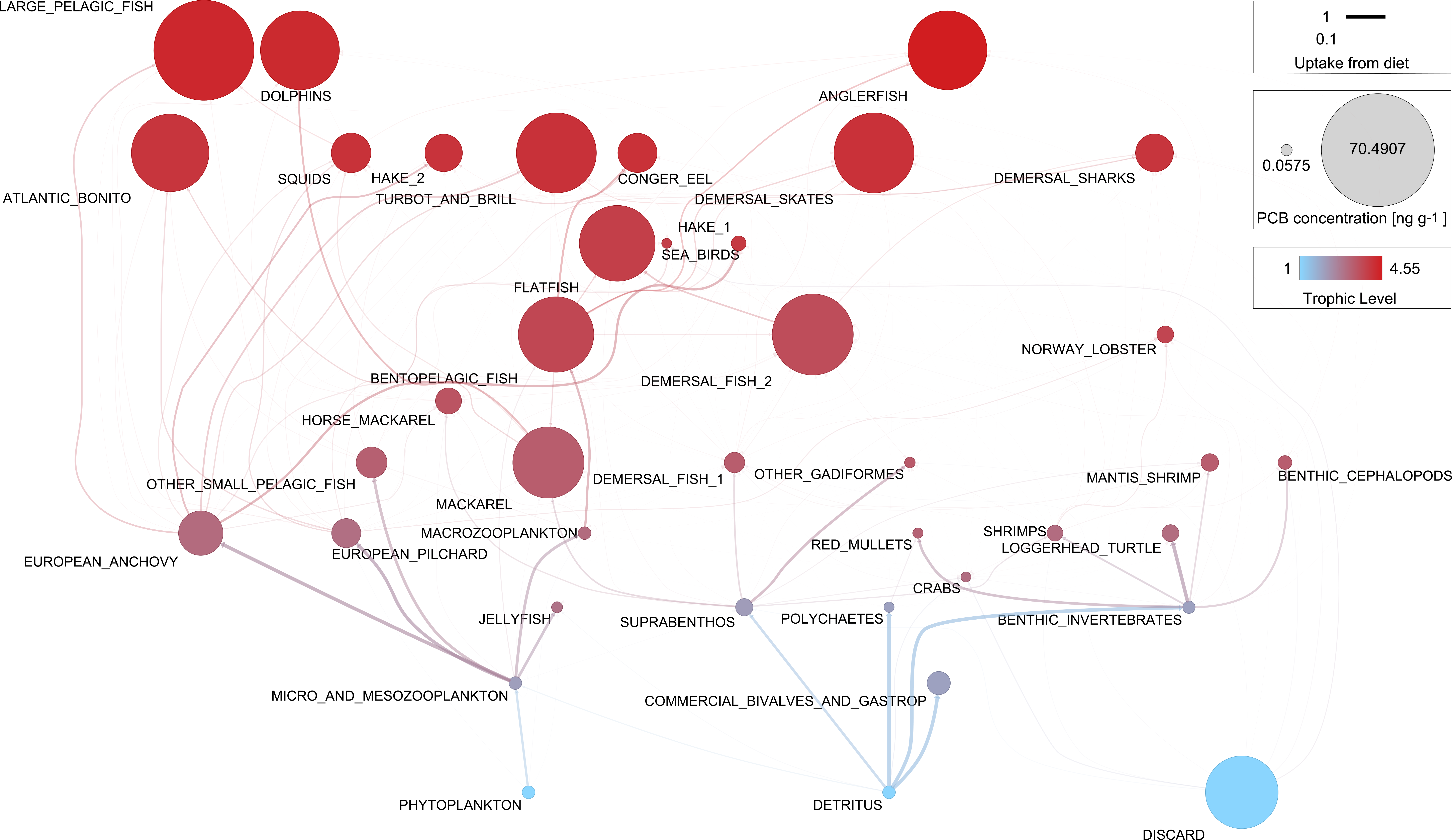}}
\caption{Estimated Adriatic trophic (a) and PCB bioaccumulation (b) networks. Nodes represent functional groups whose size is proportional to the biomass content (a) or PCB bioconcentration (b). Edges represent feeding connection and their thickness indicates the contribution of the source node in the diet of the target node (a), or in the uptake of contaminant from food in the case of bioaccumulation network (b). Flows to detritus and to fishing discards are not shown.}\label{fig:estimated_net}
\end{figure*}

The estimated trophic and contaminant flows of the Adriatic Sea model are depicted in Figure~\ref{fig:estimated_net}, and the corresponding numerical results are reported in Table \ref{tbl:biomassInOut} for the biomass network, and in Table \ref{tbl:pcb_est} for the PCBs bioaccumulation network. Estimating the trophic network with LIM required to take an approximate solution (with error tolerance $\leq 10^{-8}$) because the large amount of input trophic data taken from \citep{coll2007ecological} (reported in Table \ref{tbl:biomassInOut}) generates a high number of constraints and in turn an empty solution space. On the other hand, the partial availability of PCBs bioconcentration data produces an under-determined problem (i.e.\ many possible solutions), solved in this case through Markov Chain Monte Carlo (MCMC) sampling (5000 iterations) of the solution space.

\paragraph{Trophic Network Analysis} The analysis of the biomass network shows that species at lower trophic levels are the most prominent in terms of internal and exchanged biomass and that biomass content and trophic level are negatively correlated, as visible in Fig.\ \ref{fig:estimated_net} (a). The only exception is the \textit{Discard} group, which accounts for the discarded catches that enter back the biomass cycle. \textit{Discard} is considered in this model as a detritus, and thus has associated a trophic level (TL) of 1. However, it clearly possesses a much lower biomass than the natural detritus (group \textit{Detritus}) and primary producers (group \textit{Phytoplankton}). We report that our quantitative estimations agree with the original work by Coll et al., which allows us to validate our trophic model.


\paragraph{PCBs Bioaccumulation Analysis} On the other hand, PCBs bioaccumulation values tend to increase at higher trophic levels, thus a phenomenon of \textit{biomagnification} is clearly detectable, as one can observe in the network plot (Fig.\ \ref{fig:estimated_net} (b)).
We also evince that functional groups with high concentrations have associated low biomass values. 

We indeed register the most prominent PCBs values in top predators: \textit{Large Pelagic Fish} (TL = 4.343, PCB = 70.491), \textit{Demersal skates} (TL = 4.154, PCB = 54.833), \textit{Turbot and brill} (TL = 4.152, PCB = 54.746), \textit{Dolphins} (TL=4.302, PCB=54.048), \textit{Anglerfish} (TL = 4.553, PCB = 53.808), and \textit{Atlantic bonito} (TL = 4.087, PCB = 52.704). In particular, \textit{Large Pelagic Fish} (Tuna and Swordfish) shows by far the highest PCBs bioaccumulation, which can be explained by the concentration in groups composing its diet (mainly \textit{European Anchovy} and \textit{Squids}).

Anyway, by the nature of the LIM approach, we note that input concentration data strongly influence estimated values. Specifically, low upper bounds on PCBs values limit the output concentrations in some top predators (\textit{Squids}, \textit{Hake 2} and \textit{Demersal Sharks}). On the contrary, setting high or infinite upper bounds results in large bioaccumulation values also for groups at TL=3, like in \textit{Demersal fish 2}, \textit{Flatfish}, \textit{Bentopelagic fish} and \textit{Mackarel}. Naturally, having employed a stochastic search algorithm, the concentration variables with less constraints typically have a higher variability (see standard deviation $\sigma$ in Table \ref{tbl:pcb_est}).

Fishing activity and overexploitation represent a biomass loss, and therefore can also affect the patterns of PCBs diffusion in the ecosystem. Bioaccumulation results from the continuous uptake of pollutants over the years. Thus, fishing activity can ideally interrupt the bioaccumulation process by increasing the mortality rate of a species, even if overexploitation is not clearly an ecologically sustainable solution \citep{coll2008ecosystem}. In our model, relatively low PCBs values can be detected in correspondence of exploited species (i.e.\ with fishing rates exceeding biomass). For instance, \textit{Crabs}, \textit{Other gadiformes} and \textit{Red mullets} show concentrations substantially lower ($< 1 \ ng \cdot g^{-1}$) than those in species belonging to the same TL, but not affected by fishing pressure. A similar phenomenon is observable in \textit{Conger eel} where, albeit with a wide PCBs input range, the estimated bioaccumulation value stands close to the lower bound. This is even more evident by looking at the variations in PCBs bioaccumulation between groups describing the same species but subject to different fishing pressures, like between \textit{Hake 1} ($< 40 \ cm$, vulnerable to fishing, PCB=3.852) and \textit{Hake 2} ($> 40 \ cm$, not vulnerable, PCB=21.658); or between \textit{Demersal fish 1} (overexploited, PCB=8.159) and \textit{Demersal fish 2} (PCB=55.424). 

Differently from natural detritus, fishing discards are characterized by a significant PCBs concentration. This has to be attributed mainly to its low total biomass, combined with its species composition, that contributes to a considerable total inflow of contaminant as detailed in Table \ref{tbl:pcb_est} (column $Dis$).

Finally, with our bioaccumulation model we are also able to estimate PCBs concentration in the landing fraction of biomass exported by fishing. This is simply computed as the sum of contaminant outflows to the landings over the sum of biomass exported to the same compartment: $\dfrac{\sum_i c_{i\rightarrow Land}}{\sum_i b_{i\rightarrow Land}}$. The mean concentration value in landings equals to 18.17 $ng \cdot g^{-1}$. This kind of analysis could provide an effective indicator of the chemical pollution in species of commercial interest.

\subsection{ODE bioaccumulation model}
\begin{figure}
\centering
\includegraphics[width=0.65\columnwidth]{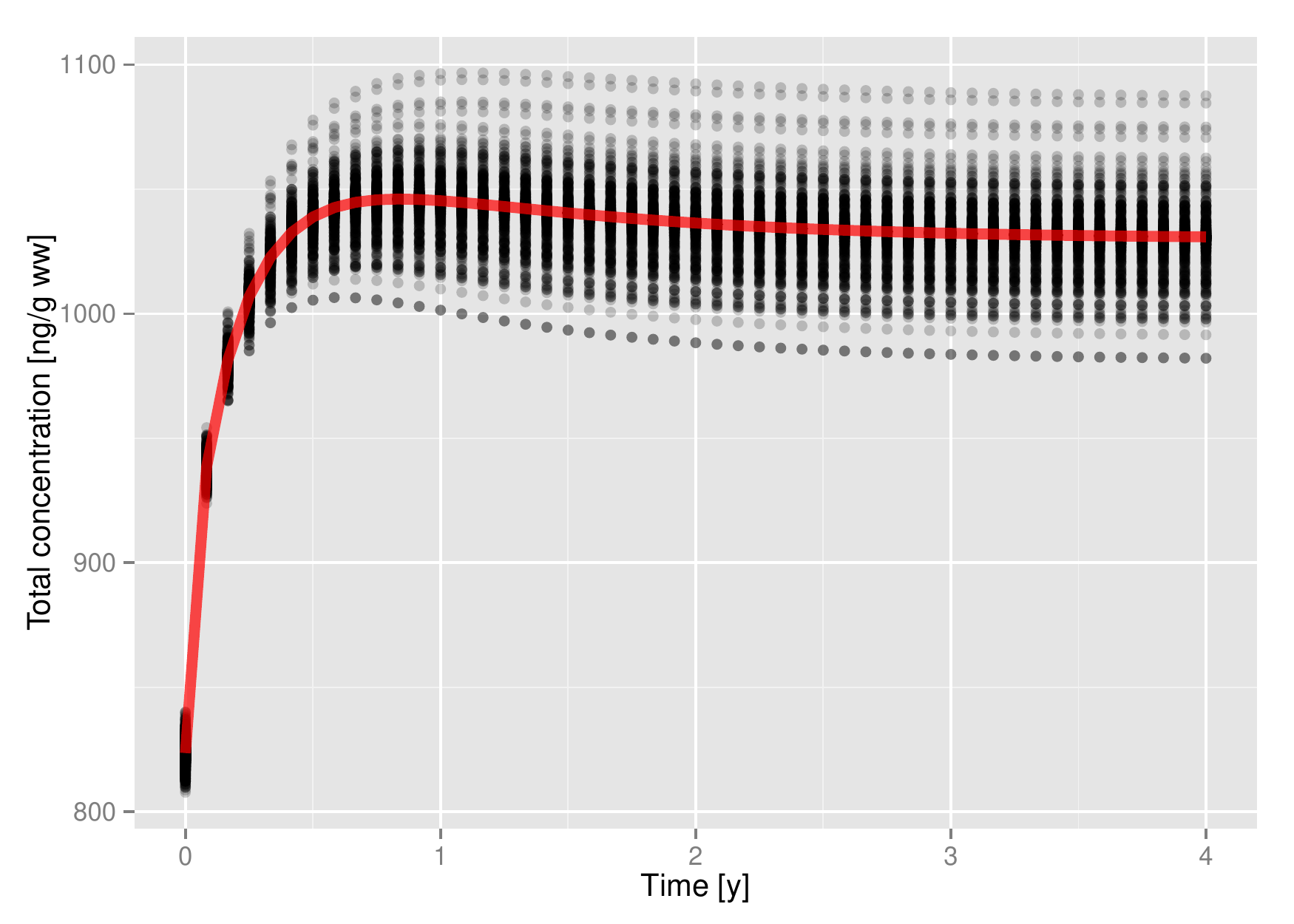}
\caption{Temporal evolution of total PCBs concentrations, simulated through the dynamic ODE model over a period of 4 years and time step of 1 month (red line). Shaded dots indicate the results obtained after 100 random perturbations of the initial concentration, for each species.}\label{fig:total_concs_ODE}
\end{figure}

We evaluate the long-term bioaccumulation dynamics in the Adriatic food web by simulating the ODE model at Eq.\ \ref{eq:lv_cont_3} for a period of 4 years with a time step of 1 month. We just report the sum of PCBs concentrations (Figure \ref{fig:total_concs_ODE}) under default initial conditions (derived from network estimations) and under random perturbations of the input concentrations (100 uniformly distributed values for each species).

We notice that the qualitative dynamics of the system are robust with respect to changes in the initial conditions. At the same time, the model is able to reproduce the quantitative impact of perturbations in PCBs concentrations, which tends to be amplified over the time. The initial steep increase in total bioaccumulation is attributable to the fact that groups in rapid equilibrium with the water compartment ($INST$ class) are not mass-balanced in the bioaccumulation network. This implies that the initial ODE state is far from equilibrium conditions, that are in any case practically reached within the first 2 years, and despite of random perturbations.

\subsection{Implementation}
Our bioaccumulation model has been implemented in R. The \textit{LIM} package \citep{van2010quantifying} was used to estimate trophic and bioaccumulation networks. This package has been preferred to the ECOPATH software \citep{christensen2004ecopath}, a de facto standard in trophic estimation and analysis, because \textit{LIM} supports custom models and equations and is general enough to describe multiple flow currencies (both biomass and PCBs), a crucial feature in our study. 

The calculation of trophic levels ($TL$ and $OI$) has been performed with the R package \textit{NetIndices}~\citep{kones2009network}, and we used the package \textit{FME} \citep{soetaert2010inverse} for ODE simulations. Network plots have been generated with \textit{Graphviz} \citep{ellson2002graphviz}. Source code is available upon request to authors.
\section{Conclusions}
The main contribution of this study is the combination of computational and network analysis tools in order to investigate
the bioaccumulation of Persistent Organic Pollutants in marine food webs.
We consider the case study of PCBs bioaccumulation in the Adriatic sea, providing a state of the art review on PCBs concentration data for the Adriatic food web and the first network level reconstruction, which allows us to evaluate the occurrence of biomagnification through trophic levels. 
In this context, Linear Inverse Modelling provides effective means to formulate our estimation problem and to deal with incomplete and uncertain data, which we quantified also with a stochastic search of the admissible contaminant flow values.

The derived dynamic ODEs simulated under random perturbation of the initial PCBs concentrations, show robust qualitative dynamics which sets the ground for a deeper study on the temporal bioaccumulation dynamics in the Adriatic ecosystem.

\bibliographystyle{model4-names}
\bibliography{lim_biorem}

\newpage

\appendix
\section{Supplementary Material}
\begin{center}
\begin{small}
\begin{longtable}{llccp{0.42\linewidth}}
\caption{Input PCBs bioconcentration data ($ng \cdot g^{-1}$) by functional group and corresponding references. In order to account for multiple data, we derive concentration ranges instead of single values.}\label{tbl:PCBinputdata}\\
\hline
Id & Group & $\sum$PCB min & $\sum$PCB max & Species and References \\ 
\hline 
1 & Phytoplankton &  &  &  \\ 
2 & Micro and mesozoop. &  &  &  \\ 
3 & Macrozooplankton &  &  &  \\ 
4 & Jellyfish &  &  &  \\ 
5 & Suprabenthos &  &  &  \\ 
6 & Polychaetes &  &  &  \\ 
7 & Commercial bivalves & 1.24 & 20.29 & \textit{M. galloprovincialis} \citep{perugini2004levels}; \textit{M. galloprovincialis, C. gallina} \citep{bayarri2001pcdds}; \textit{C. gallina, A. tubercolata, E. siliqua, M. galloprovincialis} \citep{marcotrigiano2003heavy}  \\ 
8 & Benthic Invertebrates &  &  &  \\ 
9 & Shrimps & 0.346 & 11.61 & \textit{P. longirostris, A. antennatus} \citep{marcotrigiano2003heavy}; \textit{A. antennatus, P. longirostris, P. martia} \citep{storelli2003polychlorinated}  \\ 
10 & Norway lobster & 0.2 & 10.63 & \textit{N. norvegicus} \citep{bayarri2001pcdds,perugini2004levels,storelli2003polychlorinated}  \\ 
11 & Mantis shrimp & 2.64 & 11.61 &  \textit{S. mantis} \citep{marcotrigiano2003heavy}\\ 
12 & Crabs &  &  &  \\ 
13 & Benthic cephalopods & 0.31 & 6.7 & \textit{T. sagittatus, S. officinalis} \citep{perugini2004levels}; \textit{O. salutii} \citep{marcotrigiano2003heavy}  \\ 
14 & Squids & 9.53 & 37.7 & \textit{L. vulgaris} (bayarri2001pcdds); \textit{I. coindetii} \citep{marcotrigiano2003heavy} \\ 
15 & Hake 1 & \multirow{2}{*}{3.183} & \multirow{2}{*}{31.93} & \multirow{2}{*}{\parbox{0.98\linewidth}{\textit{M. merluccius} \citep{storelli2003polychlorinated, marcotrigiano2003heavy}}} \\ 
16 & Hake 2 & & &\\ 
17 & Other gadiformes &  &  &  \\ 
18 & Red mullets &  &  &  \\ 
19 & Conger eel & 22.424 & 104 &\textit{C. conger} \citep{storelli2003polychlorinated, storelli2007metals}\\ 
20 & Anglerfish & 0.2 &  & \textit{L. boudegassa} \citep{storelli2003polychlorinated} \\ 
21 & Flatfish &  &  &  \\ 
22 & Turbot and brill &  &  &  \\ 
23 & Demersal sharks & 2 & 42 & \textit{C. granolousus, S. blainvillei} \citep{storelli2001persistent}; \\ 
24 & Demersal skates & 0.45 &  & \textit{R. miraletus, R. clavata, R. oxyrhincus} \citep{storelli2003polychlorinated}; \\ 
25 & Demersal fish 1 & 6.687 &  & \multirow{3}{0.42\columnwidth}{\textit{S. flexuosa, H. dactyloptereus} \citep{storelli2003polychlorinated}}\\ 
26 & Demersal fish 2 & 6.687 &  &  \\ 
27 & Bentopelagic fish & 6.687 &  &  \\ 
28 & European Anchovy & 1.22 & 62.7 & \textit{E. encrasicolus} \citep{perugini2004levels, bayarri2001pcdds} \\ 
29 & European Pilchard & 5.327 & 26.25 & \textit{S. pilchardus} \citep{perugini2004levels, storelli2003polychlorinated} \\ 
30 & Small Pelagic Fish & 4.54 & 31.9 & \textit{S. aurita} \citep{marcotrigiano2003heavy}; \textit{S. maris, A. rochei} \citep{storelli2003polychlorinated}  \\ 
31 & Horse Mackarel & 6.761 & & \textit{T. trachurus} \citep{storelli2003polychlorinated} \\ 
32 & Mackarel & 0.95 & 80.6 & \textit{S. scombrus} \citep{perugini2004levels,bayarri2001pcdds} \\ 
33 & Atlantic bonito &  &  &  \\ 
34 & Large Pelagic Fish &  &  &  \\ 
35 & Dolphins &  &  &  \\ 
36 & Loggerhead turtle & 0.63 & 23.49 & \textit{C. caretta} \citep{storelli2007polychlorinated, corsolini2000presence} \\ 
37 & Sea birds &  &  &  \\ 
38 & Discard &  &  &  \\ 
39 & Detritus &  &  &  \\
\hline
\end{longtable}
\end{small}
\end{center}

\begin{figure*}
\centering
\subfloat[Diet composition matrix]{\includegraphics[width=0.8\linewidth]{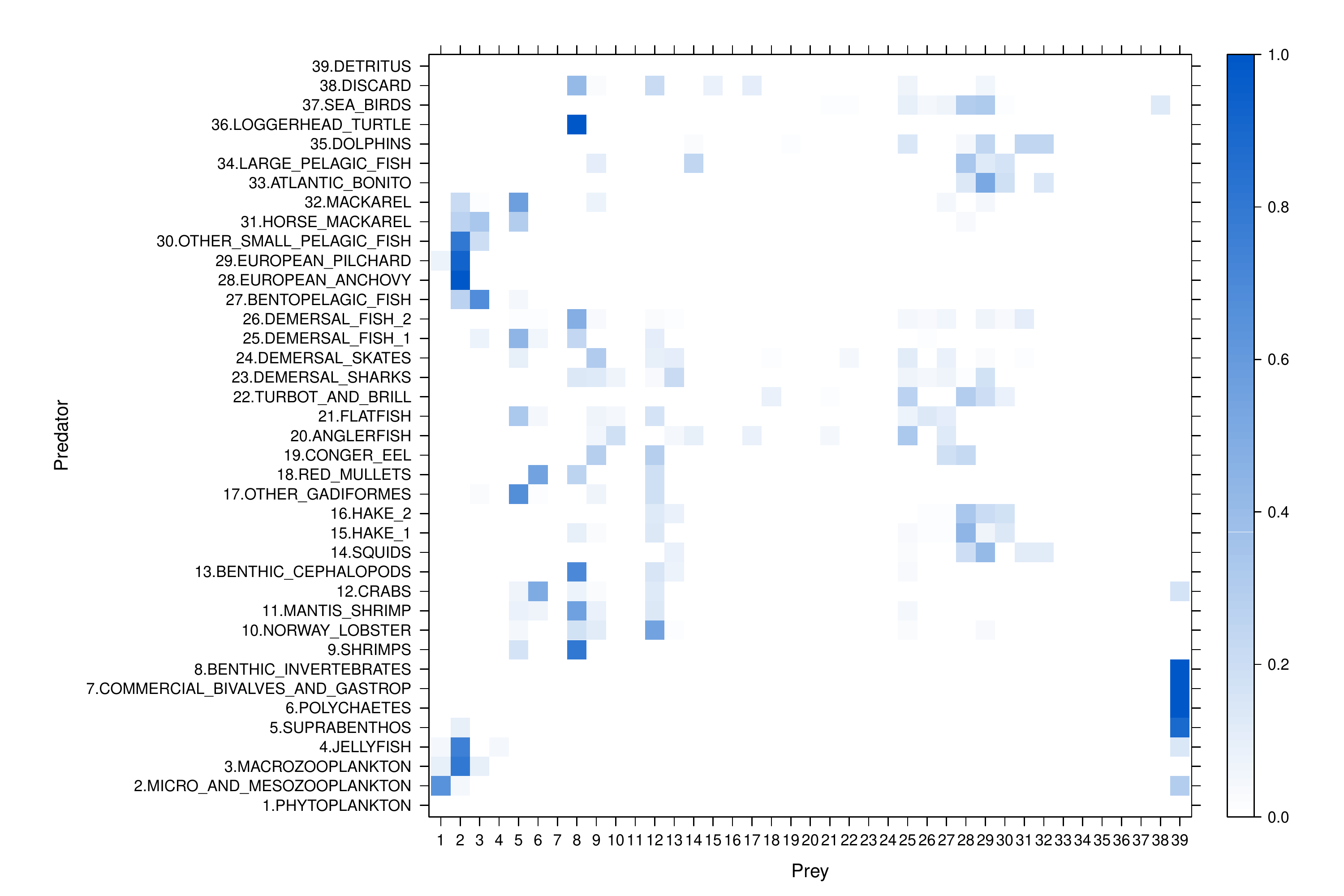}}\\
\subfloat[Contaminant uptake from diet matrix]{\includegraphics[width=0.8\linewidth]{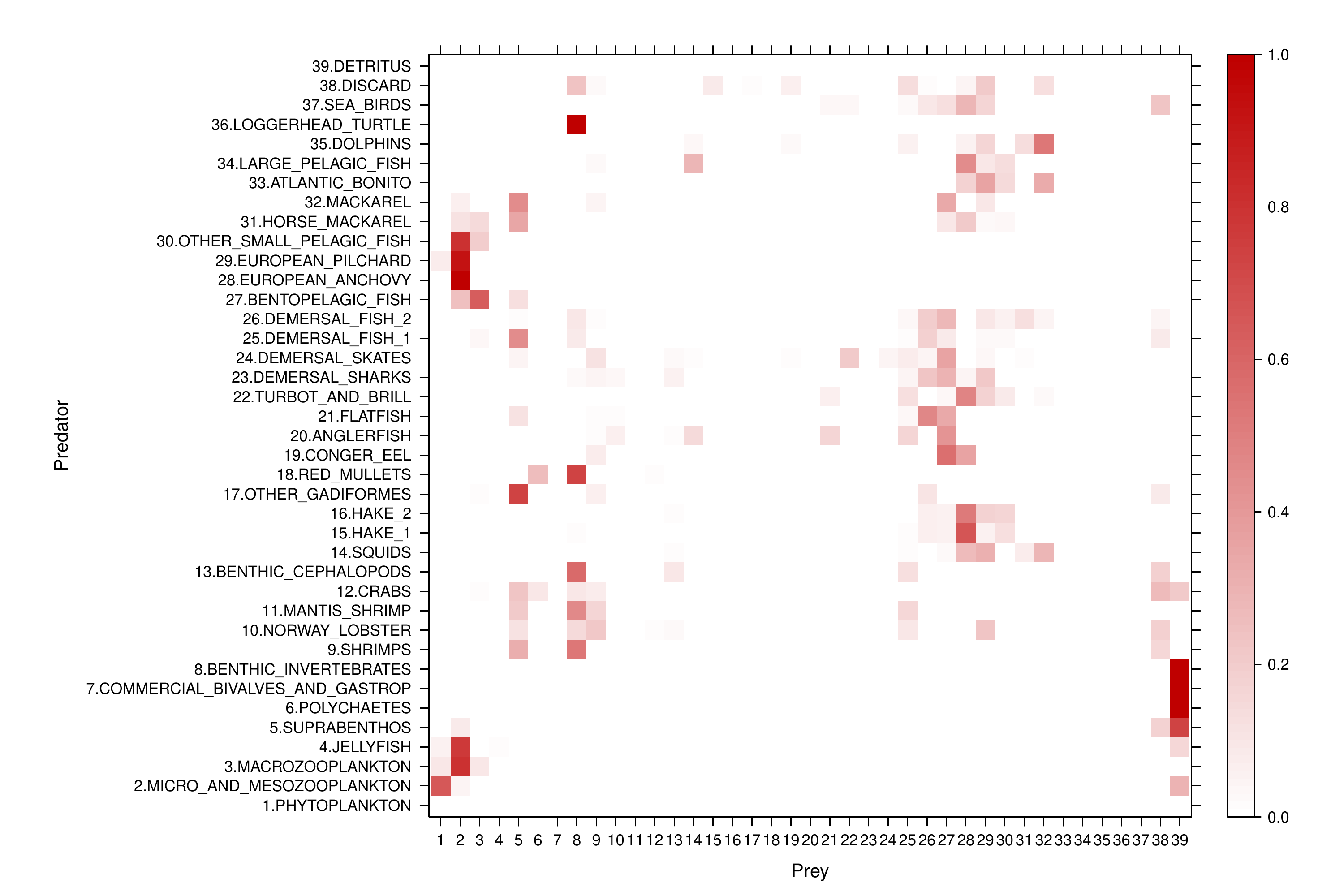}}
\caption{Level plots of the diet composition matrix in the trophic network (a) and of the contaminant uptake rate from diet relative to the PCB bioaccumulation network. Darker cells indicate feeding links where the contribution of the prey in the diet/PCBs concentration of the predator is higher. Diet composition has been taken from \cite{coll2007ecological}, while the uptake rate of a predator $j$ from a prey $i$, $U_{ij}$, is the contaminant flow from $i$ to $j$ scaled by the sum of the inflows of $j$.}\label{fig:dietcompositions}
\end{figure*}

\begin{landscape}
\label{tbl:biomassInOut}
\centering
\begin{small}
\begin{longtable}{lllll|llllll}
\caption{Input and estimated data for the trophic newtork. Input data are: $B_\text{in}$, input biomass ($t\cdot km^{-2}$); $P$, production rate ($yr^{-1}$); $Q$, consumption rate ($yr^{-1}$). $B_\text{out}$ ($t\cdot km^{-2}$) is the estimated biomass; $F$ is the biomass exports due to fishing ($t\cdot km^{-2} \cdot yr^{-1}$), and $Dis$ the fraction discarded. Trophic level analysis is summarized in $OI$, omnivory index; and $TL$, trophic level.}\\

\hline
Id & Group & $B_\text{in}$ & $P$ & $Q$ & $B_\text{out}$ & $F$ & $Dis$ & $OI$ & $TL$\\  \hline
1 & Phytoplankton & 16.658 & 69.03 &  & 16.658 & 0 & 0 & 0 & 1\\ 
2 & Micro and mesozooplankton & 9.512 & 30.43 & 49.87 & 9.512 & 0 & 0 & 0.053 & 2.053\\ 
3 & Macrozooplankton & 0.54 & 21.28 & 53.14 & 0.54 & 0 & 0 & 0.210 & 3.047\\ 
4 & Jellyfish & 4 & 14.6 & 50.48 & 4 & 0 & 0 & 0.228 & 2.884\\ 
5 & Suprabenthos & 1.01 & 8.4 & 54.36 & 1.01 & 0 & 0 & 0.100 & 2.105\\ 
6 & Polychaetes & 9.984 & 1.9 & 11.53 & 9.984 & 0 & 0 & 0 & 2\\ 
7 & Commercial bivalves and gastrop & 0.043 & 1.06 & 3.13 & 0.043 & 0.035 & 0 & 0 & 2\\ 
8 & Benthic Invertebrates & 79.763 & 1.06 & 3.13 & 79.763 & 0.328 & 0.328 & 0 & 2\\ 
9 & Shrimps &  & 3.21 & 7.2 & 0.68 & 0.033 & 0.017 & 0.022 & 3.018\\ 
10 & Norway lobster & 0.018 & 1.25 & 4.56 & 0.018 & 0.037 & 0 & 0.212 & 3.771\\ 
11 & Mantis shrimp & 0.015 & 1.5 & 4.56 & 0.015 & 0.072 & 0 & 0.226 & 3.307\\ 
12 & Crabs & 0.009 & 2.44 & 4.73 & 0.009 & 0.179 & 0.177 & 0.352 & 2.998\\ 
13 & Benthic cephalopods & 0.068 & 2.96 & 5.3 & 0.068 & 0.156 & 0.002 & 0.281 & 3.307\\ 
14 & Squids & 0.02 & 3.11 & 26.47 & 0.02 & 0.041 & 0 & 0.040 & 4.140\\ 
15 & Hake 1 & 0.06 & 1 & 4.24 & 0.06 & 0.183 & 0.07 & 0.132 & 3.996\\ 
16 & Hake 2 &  & 0.5 & 1.85 & 0.56 & 0 & 0 & 0.027 & 4.114\\ 
17 & Other gadiformes & 0.029 & 1.59 & 4.37 & 0.029 & 0.108 & 0.083 & 0.202 & 3.369\\ 
18 & Red mullets & 0.025 & 1.9 & 8.02 & 0.025 & 0.112 & 0 & 0.153 & 3.190\\ 
19 & Conger eel & 0.005 & 1.92 & 6.45 & 0.005 & 0.008 & 0.008 & 0.078 & 4.156\\ 
20 & Anglerfish & 0.006 & 1.04 & 4.58 & 0.006 & 0.007 & 0 & 0.095 & 4.553\\ 
21 & Flatfish & 0.009 & 1.43 & 9.83 & 0.009 & 0.04 & 0 & 0.451 & 3.886\\ 
22 & Turbot and brill &  & 1.43 & 5.34 & 0.04 & 0.016 & 0 & 0.046 & 4.152\\ 
23 & Demersal sharks & 0.018 & 0.63 & 4.47 & 0.018 & 0.008 & 0 & 0.260 & 4.086\\ 
24 & Demersal skates & 0.003 & 1.11 & 7.08 & 0.003 & 0.002 & 0 & 0.252 & 4.154\\ 
25 & Demersal fish 1 & 0.056 & 2.4 & 7.68 & 0.056 & 0.106 & 0.051 & 0.238 & 3.315\\ 
26 & Demersal fish 2 &  & 2.4 & 5.68 & 0.24 & 0.017 & 0.001 & 0.495 & 3.619\\ 
27 & Bentopelagic fish &  & 1.07 & 7.99 & 1.2 & 0.002 & 0 & 0.212 & 3.731\\ 
28 & European Anchovy & 1.019 - 6.611 & 0.87 & 11.02 & 1.497 & 0.501 & 0.005 & 0 & 3.053\\ 
29 & European Pilchard & 2.985 - 7.803 & 0.75 & 9.19 & 2.985 & 0.406 & 0.042 & 0.082 & 2.968\\ 
30 & Other small Pelagic Fish & 0.413 - 1.517 & 1.1 & 11.29 & 1.517 & 0.013 & 0.001 & 0.158 & 3.251\\ 
31 & Horse Mackarel & 0.659 - 2.455 & 0.99 & 7.57 & 2.455 & 0.022 & 0.002 & 0.243 & 3.493\\ 
32 & Mackarel & 0.452 - 1.683 & 0.99 & 6.09 & 1.683 & 0.025 & 0.008 & 0.217 & 3.319\\ 
33 & Atlantic bonito & 0.3 & 0.39 & 4.54 & 0.3 & 0.018 & 0 & 0.021 & 4.087\\ 
34 & Large Pelagic Fish & 0.138 & 0.37 & 1.99 & 0.138 & 0.026 & 0 & 0.219 & 4.343\\ 
35 & Dolphins & 0.012 & 0.08 & 11.01 & 0.012 & 0.0001 & 0.0001 & 0.076 & 4.302\\ 
36 & Loggerhead turtle & 0.032 & 0.17 & 2.54 & 0.032 & 0.004 & 0.004 & 0.010 & 3.010\\ 
37 & Sea birds & 0.001 & 4.61 & 69.34 & 0.001 & 0 & 0 & 0.597 & 3.899\\ 
38 & Discard & 0.733 &  &  & 0.733 & 0 & 0 & 0 & 1\\ 
39 & Detritus & 200 &  &  & 200 & 0 & 0 & 0 & 1\\ \hline
\end{longtable}
\end{small}
\end{landscape}

\begin{table}
\caption{PCBs concentrations ($\sum$PCB$_\text{est}$, $ng \cdot g^{-1}$) and standard deviations ($\sigma$, $ng \cdot g^{-1}$) obtained with MCMC-based sampling. $F$ ($t \cdot ng \cdot y^{-1} \cdot km^{-2}$) indicates PCBs outflows due to fishing, and $Dis$ the fraction discarded.}\label{tbl:pcb_est}
\centering
\begin{small}
\begin{tabular}{llllll}
\hline
Id & Group & $\sum$PCB$_\text{est}$ & $\sigma$ & $F$ & $Dis$ \\ \hline
1 & Phytoplankton & 2.247 & 0.955 & 0 & 0\\ 
2 & Micro and mesozooplankton & 2.247 & 0.955 & 0 & 0\\ 
3 & Macrozooplankton & 2.247 & 0.955 & 0 & 0\\ 
4 & Jellyfish & 0.843 & 0.353 & 0 & 0\\ 
5 & Suprabenthos & 5.952 & 3.247 & 0 & 0\\ 
6 & Polychaetes & 0.348 & 0.200 & 0 & 0\\ 
7 & Commercial bivalves and gastrop & 10.502 & 5.369 & 0.368 & 0\\ 
8 & Benthic Invertebrates & 2.120 & 0.846 & 0.696 & 0.696\\ 
9 & Shrimps & 4.746 & 2.752 & 0.157 & 0.081\\ 
10 & Norway lobster & 5.542 & 3.115 & 0.205 & 0\\ 
11 & Mantis shrimp & 6.103 & 3.228 & 0.439 & 0\\ 
12 & Crabs & 0.058 & 0.039 & 0.010 & 0.010\\ 
13 & Benthic cephalopods & 3.198 & 1.770 & 0.499 & 0.006\\ 
14 & Squids & 23.289 & 8.172 & 0.955 & 0\\ 
15 & Hake 1 & 3.852 & 0.617 & 0.705 & 0.270\\ 
16 & Hake 2 & 21.658 & 5.950 & 0 & 0\\ 
17 & Other gadiformes & 0.567 & 0.285 & 0.061 & 0.047\\ 
18 & Red mullets & 0.385 & 0.262 & 0.043 & 0\\ 
19 & Conger eel & 22.879 & 0.443 & 0.183 & 0.183\\ 
20 & Anglerfish & 53.808 & 29.234 & 0.377 & 0\\ 
21 & Flatfish & 51.436 & 30.137 & 2.057 & 0\\ 
22 & Turbot and brill & 54.746 & 28.303 & 0.876 & 0\\ 
23 & Demersal sharks & 21.840 & 11.571 & 0.175 & 0\\ 
24 & Demersal skates & 54.833 & 29.337 & 0.110 & 0\\ 
25 & Demersal fish 1 & 8.159 & 1.132 & 0.865 & 0.416\\ 
26 & Demersal fish 2 & 55.424 & 28.667 & 0.942 & 0.055\\ 
27 & Bentopelagic fish & 51.324 & 28.892 & 0.103 & 0\\ 
28 & European Anchovy & 27.104 & 18.170 & 13.579 & 0.136\\ 
29 & European Pilchard & 14.969 & 6.234 & 6.077 & 0.629\\ 
30 & Other small Pelagic Fish & 16.533 & 8.180 & 0.215 & 0.017\\ 
31 & Horse Mackarel & 12.496 & 1.171 & 0.275 & 0.025\\ 
32 & Mackarel & 47.960 & 21.422 & 1.199 & 0.384\\ 
33 & Atlantic bonito & 52.704 & 29.504 & 0.949 & 0\\ 
34 & Large Pelagic Fish & 70.491 & 20.461 & 1.833 & 0\\ 
35 & Dolphins & 54.048 & 29.286 & 0.005 & 0.005\\ 
36 & Loggerhead turtle & 5.478 & 4.412 & 0.022 & 0.022\\ 
37 & Sea birds & 0.161 & 0.058 & 0 & 0\\ 
38 & Discard & 49.005 & 0.840 & 0 & 0\\ 
39 & Detritus & 2.247 & 0.955 & 0 & 0\\ \hline
\end{tabular}
\end{small}
\end{table}

\begin{landscape}
\centering
\label{tbl:detailedPCBrefs}
\begin{small}
\begin{longtable}{p{0.1\linewidth}lp{0.08\linewidth}p{0.3\linewidth}llp{0.15\linewidth}}
\caption{Summary of the studies considered for the parametrization of the PCBs bioaccumulation model. We report reference; period and area of analysis; species considered; kind of tissue sampled; units of measurement; and PCBs congeners detected.}\\
\hline
Reference & Period & Area & Species & Tissue & Units & PCBs congeners\\ \hline
\citep{perugini2004levels} & 2002 & Center  & \textit{M. galloprovincialis,  N. norvegicus, M. barbatus, S. officinalis, E. flying squid,  E. encrasicholus,   S. pilchardus,  S.r scombrus} & edible & ng/g ww  & 28, 52, 101, 118, 138, 153, 180\\ 
\citep{marcotrigiano2003heavy} &  & Adriatic, Ionian Sea & \textit{M. Merluccius, M. poutassou, P. blennoides, S. smaris, S. pilchardus, E. encrasicholus, L. caudatus, H. dactylopterus, L. budegassa, T. trachurus, A. rochei,   Raje spp,  P. glauca, S. acanthias, S. blainvillei, S. canicula, G. melastomus, C.  gallina, A.  tubercolata,  E.  siliqua, M. galloprovincialis, O. salutii,  I. coindeti,  S. mantis, P. longirostris, A.  antennatus}  & muscle  & ng/g ww  & 8, 20, 28, 35, 52, 60, 77, 101, 105, 118, 126, 138, 153, 156, 169, 180, 209\\ 
\citep{bayarri2001pcdds} & 1997-1998 & North, Center, South & \textit{L. vulgaris, M. galloprovincialis, N. norvegicus, M. barbatus, C. gallina} & edible & ng/g ww  & 28, 52, 101, 118, 138, 153, 180, 138, 163 \\ 
\citep{storelli2003polychlorinated} & 2001 & South & \textit{C. conger, H. dactylopterus, L. boudegassa, M. barbatus, S. flexuosa, P. blennoides, P. erythrinus, R. clavata, R. oxyrinchus, R. miraletus, S. pilchardus, M. merluccius, S. aurita, S. scombrus, T. trachurus, A. antennatus, N. norvegicus, P. longirostris, P.  martia} & muscle  & pg/g ww & 60, 77, 101, 105, 118, 126, 138, 153, 156, 169, 180, 209\\ 
\citep{storelli2007metals} &  & South & \textit{A. anguilla} & muscle  & ng/g ww  & 52, 70, 77, 101, 105, 118, 126, 138, 153, 180\\ 
\citep{storelli2001persistent} & 1999 & South & \textit{C. granulosus, S. blainvillei} & muscle  & ng/g ww  & 8, 20, 28, 35, 52, 60, 77, 101, 105, 118, 126, 138, 153, 156, 169, 180, 209 \\ 
\citep{storelli2007polychlorinated} &  & Adriatic, Ionian Sea & \textit{C. caretta} & muscle  & ng/g ww  &  8, 20, 28, 35, 52, 60, 77, 101, 105, 118, 126, 138, 153, 156, 169, 180, 209\\ 
\citep{corsolini2000presence} & 1994 & Adriatic, Baltic, Northern Sea & \textit{C. caretta} & muscle  & ng/g ww  & 153, 137, 138, 180, 170, 194, 60, 118, 105, 156, 189, 77, 126, 169\\
\hline
\end{longtable}
\end{small}
\end{landscape}







\end{document}